\documentclass[reprint,aps,prl,superscriptaddress,showpacs,footinbib,longbibliography,amsmath]{revtex4-1}
\usepackage{graphicx}
\usepackage{dcolumn}
\usepackage{bm}

\graphicspath{{../}}

\usepackage{soul}

\usepackage{color}
\usepackage[colorlinks,bookmarks=false,citecolor=darkblue,linkcolor=red,urlcolor=blue]{hyperref}

\definecolor{darkred}{rgb}{0.7,0.0,0.0}

\definecolor{darkblue}{rgb}{0,0.02,0.45}

\definecolor{darkgreen}{rgb}{0.02,0.45,0.0}

\definecolor{violet}{rgb}{0.8,0.2,0.6}

\begin{document}
\title{Field evolution of low-energy excitations in the hyperhoneycomb magnet $\beta$-Li$_2$IrO$_3$}
\author{M. Majumder}
\email{mayukh.cu@gmail.com}
\affiliation{Experimental Physics VI, Center for Electronic Correlations and Magnetism, University of Augsburg, 86159 Augsburg, Germany}

\author{M. Prinz-Zwick}
\author{S. Reschke}
\affiliation{Experimental Physics V, Center for Electronic Correlations and Magnetism, University of Augsburg, 86159 Augsburg, Germany}

\author{A. Zubtsovskii}
\author{T. Dey}
\author{F.~Freund}
\affiliation{Experimental Physics VI, Center for Electronic Correlations and Magnetism, University of Augsburg, 86159 Augsburg, Germany}

\author{N. B{\"u}ttgen}
\affiliation{Experimental Physics V, Center for Electronic Correlations and Magnetism, University of Augsburg, 86159 Augsburg, Germany}

\author{A. Jesche}
\affiliation{Experimental Physics VI, Center for Electronic Correlations and Magnetism, University of Augsburg, 86159 Augsburg, Germany}

\author{I. K\'{e}zsm\'{a}rki}
\affiliation{Experimental Physics V, Center for Electronic Correlations and Magnetism, University of Augsburg, 86159 Augsburg, Germany}

\author{A. A. Tsirlin}
\email{altsirlin@gmail.com}
\author{P. Gegenwart}
\affiliation{Experimental Physics VI, Center for Electronic Correlations and Magnetism, University of Augsburg, 86159 Augsburg, Germany}

\date{\today}

\begin{abstract}
$^7$Li nuclear magnetic resonance (NMR) and terahertz (THz) spectroscopies are used to probe magnetic excitations and their field dependence in the hyperhoneycomb Kitaev magnet $\beta$-Li$_2$IrO$_3$. Spin-lattice relaxation rate ($1/T_1$) measured down to 100\,mK indicates gapless nature of the excitations at low fields (below $H_c\simeq 2.8$\,T), in contrast to the gapped magnon excitations found in the honeycomb
Kitaev magnet $\alpha$-RuCl$_3$ at zero applied magnetic field. At higher temperatures in $\beta$-Li$_2$IrO$_3$, $1/T_1$ passes through a broad maximum without any clear anomaly at the N\'eel temperature $T_N\simeq 38$\,K, suggesting the abundance of low-energy excitations that are indeed observed as two peaks in the THz spectra, both correspond to zone-center magnon excitations. At higher fields (above $H_c$), an excitation gap opens, and a re-distribution of the THz spectral weight is observed without any indication of an excitation continuum, in contrast to $\alpha$-RuCl$_3$ where an excitation continuum was reported.
\end{abstract}

\maketitle
\textit{Introduction.}
Materials with strong spin-orbit coupling host highly anisotropic exchange interactions that cause unusual magnetically ordered states and exotic excitations. This physics is currently under active investigation in honeycomb and honeycomb-like materials with bond-dependent Kitaev interactions~\cite{hermanns2018}. The pure Kitaev model features a spin-liquid ground state characterized by fractionalized excitations~\cite{knolle2014}, but the majority of materials reported to date are magnetically ordered in zero field, because additional interactions beyond the Kitaev term are present~\cite{winter2017r}. Nevertheless, at least one of these materials, $\alpha$-RuCl$_3$, reveals not only magnon excitations at low energies~\cite{ran2017,wu2018,shi2018,ozel2019}, but also a peculiar excitation continuum at higher energies~\cite{banerjee2017,wang2017,wellm2018,reschke2019}. This continuum is sometimes interpreted as a vestige of spin-liquid physics of the pure Kitaev model~\cite{banerjee2018}, although a more mundane explanation in terms of magnon breakdown caused by anisotropic terms in the spin Hamiltonian appears equally plausible and even better justified microscopically~\cite{winter2017,winter2018}.

With the exception of $\alpha$-RuCl$_3$, Kitaev materials are based on iridium~\cite{winter2017r} and, thus, notoriously difficult for neutron-scattering studies. Consequently, little information on their magnetic excitations is available~\cite{choi2012,choi2019}. Raman scattering was used to observe broad excitation continua in Na$_2$IrO$_3$ and $\alpha$-Li$_2$IrO$_3$~\cite{gupta2016}. The temperature evolution of this spectral feature may be indicative of fractionalized excitations of the Kitaev model~\cite{nasu2016}. 

Here, we focus on magnetic excitations of the hyperhoneycomb Kitaev iridate $\beta$-Li$_2$IrO$_3$~\cite{biffin2014,takayama2015,majumder2019}. This compound is magnetically ordered at ambient pressure and in zero magnetic field, but its incommensurately ordered state appears to be rather fragile. Pressure breaks down the magnetic order and triggers the formation of a partially frozen spin liquid above 1.4\,GPa~\cite{majumder2018}, which may be concomitant with a structural transformation~\cite{veiga2019}. External field applied along the $b$ direction alters the magnetically ordered state too, although in this case the incommensurate order is gradually replaced by the commensurate one above $H_c=2.8$\,T~\cite{ruiz2017,rousochatzakis2018,ducatman2018}, and no spin-liquid state is observed. Raman scattering revealed a broad excitation continuum centered around 30\,meV at ambient pressure~\cite{glamazda2016}, a remarkably high energy, given that leading exchange couplings in this compound do not exceed $10-15$\,meV~\cite{katukuri2016,majumder2018,majumder2019}. 

In the following, we use $^7$Li nuclear magnetic resonance (NMR) and terahertz (THz) spectroscopy measurements to probe excitations that lie below this putative continuum and manifest themselves by an unusual temperature dependence of the spin-lattice relaxation rate. Field evolution of these excitations is also peculiar and follows the evolution of magnetic order as a function of field. 

\textit{Samples.} Single crystals of $\beta$-Li$_2$IrO$_3$ are less than 0.5\,mm in size~\cite{biffin2014,ruiz2017,majumder2019} and too small for THz measurements. Although NMR measurements at 4.23\,T and above 15\,K could be performed on a mosaic of several co-aligned crystals~\cite{majumder2019}, this approach becomes unfeasible for lower magnetic fields and at lower temperatures that are essential to probe the presence of the magnon gap. Therefore, we used polycrystallines samples for all measurements reported in this work. The polycrystalline samples were prepared by a conventional solid-state reaction, as described in Ref.~\onlinecite{majumder2018}. Sample quality was verified by powder x-ray diffraction and magnetization measurements. In zero field, a sharp magnetic transition at $T_N\simeq 38$\,K was observed in all samples used in this study.

Polycrystalline samples contain randomly oriented crystallites and thus feature both $H\,\|\,b$ and $H\!\!\perp\!b$ configurations that lead to the coexistence of two magnetic phases, commensurate and incommensurate, above $H_c$. Although only a limited information can be obtained on polycrystalline samples in this mixed-phase regime, we are able to track several changes that occur around $H_c$, most notably, the evolution of the magnon gap across this transition.

\textit{$^7$Li Nuclear Magnetic Resonance.} The NMR measurements were performed at several different frequencies, and field-sweep spectra were taken using the conventional pulsed NMR technique at each frequency. The data were collected down to 1.8\,K and extended down to 100\,mK at several selected frequencies using a $^3$He/$^4$He dilution-fridge. 

\begin{figure}
{\centering {\includegraphics[width=10cm]{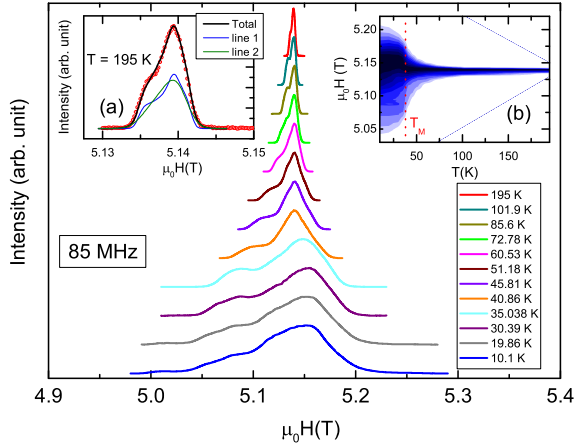}}\par} \caption{\label{fig:spectra}$^7$Li NMR spectra at different temperatures measured at a fixed frequency of 85\,MHz. Inset (a): $^7$Li NMR spectra at 195\,K at a fixed frequency of 85\,MHz along with the fitting. inset (b): Temperature evolution of the $^7$Li NMR spectra as a contour plot.} \label{structure}
\end{figure}

Temperature dependence of the $^7$Li ($I=\frac32$) NMR spectra measured at 85\,MHz is shown in Fig.~\ref{fig:spectra}. The spectra are deconvoluted into two lines assuming two Li sites with anisotropic shifts shown in the inset (a) of Fig.~\ref{fig:spectra}. Satellite lines expected for $^7$Li with $I=\frac32>\frac12$ due to the quadrupolar interaction could not been observed, likely because of the rather low electric field gradient produced by the surrounding charges at the Li sites. The areas under each Li line are found to be equal due to the equal occupancy of the crystallographically non-equivalent Li1 and Li2 sites. With decreasing temperature, the spectra become more anisotropic and complex as the anisotropic shifts of the two Li sites behave differently~\cite{majumder2019}, which hinders a reliable estimation of the shift parameters as a function of temperature. The inset (b) of Fig.~\ref{fig:spectra} clearly shows the abrupt broadening of the spectra below the ordering temperature ($T_N\approx 38$\,K), suggesting the development of local internal fields in the magnetically ordered state.

To study the field evolution of low-energy excitations, we measured the temperature dependence of the spin-lattice relaxation time ($T_1$) in a broad range of magnetic fields across the critical field $H_c$. $T_1$ has been measured at the peak frequency, where both $H\,\|\,b$ and $H\!\!\perp\!b$ crystallites contribute to the spectrum~\cite{majumder2019}, and estimated by fitting the recovery curve (obtained by the saturation recovery method) with the equation $1-M(\tau)/M(\infty) = C\times\exp[-(\tau/T_1)^\beta]$, where $M$ is the nuclear magnetization and $\beta$ is the stretching exponent. $T_1$ is related to the electron spin dynamics via the following equation
\begin{equation}
1/T_1T \propto \sum_{q,\omega_n\rightarrow 0} A_{\rm hf}^2(q)\cdot \frac{\chi''_\perp(q,\omega_n)}{\omega_n}
\end{equation}
where $\chi''(q,\omega_n)$ is the imaginary part of the dynamic spin susceptibility, $\omega_n$ the nuclear Larmor frequency, and $A_{\rm hf}(q)$ is the $q$-dependent hyperfine form-factor.

\begin{figure}
{\centering {\includegraphics[width=8.5cm]{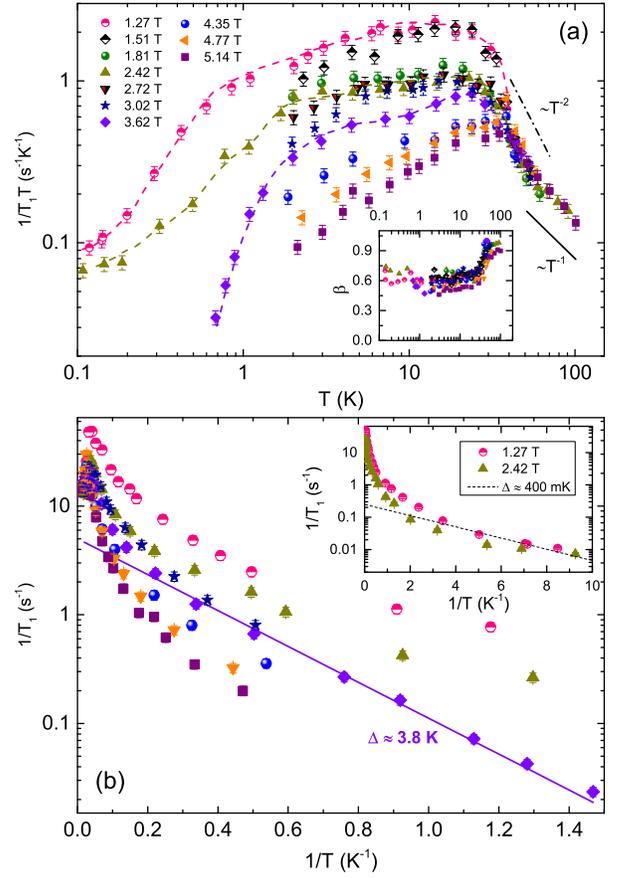}}\par} \caption{\label{fig:T1T} (a): Temperature dependence of $1/T_1T$ at different magnetic fields. Dashed lines are guide-for-the-eye. The inset shows $\beta$ as a function of temperature at different fields. (b): Arrhenius plot of $1/T_1$ versus $1/T$ for different applied magnetic fields. The straight lines are the exponential fits ($1/T_1 = A \times e^{-\Delta/T}$) to the data below 5\,K. Inset: Arrhenius plot of $1/T_1$ versus $1/T$ down to the lowest temperatures for two field values below $H_c$.}
\end{figure}

Fig.~\ref{fig:T1T}(a) presents $1/T_1T$ as a function of temperature in different applied magnetic fields. Above $T_N$, $1/T_1T$ and, hence, the spin susceptibility do not show any appreciable field dependence. For temperatures above 70\,K (well above the ordering temperature), $1/T_1T$ is inversely proportional to the temperature, i.e., $1/T_1$ = const, thus reflecting Curie susceptibility of local moments (shown by the solid line in Fig.~\ref{fig:T1T}(a)). Towards lower temperatures, $T_N<T<70$\,K, $1/T_1T$ increases steeper than the Curie law would predict (indicated by the dashed line in Fig.~\ref{fig:T1T}(a)), a signature of the critical slowing down of spin fluctuations upon approaching the ordering temperature $T_N$. For the lowest applied magnetic fields this steep increase of $1/T_1T$ is even stronger and appears right below $T_N$. 

\begin{figure}
{\centering {\includegraphics[width=9cm]{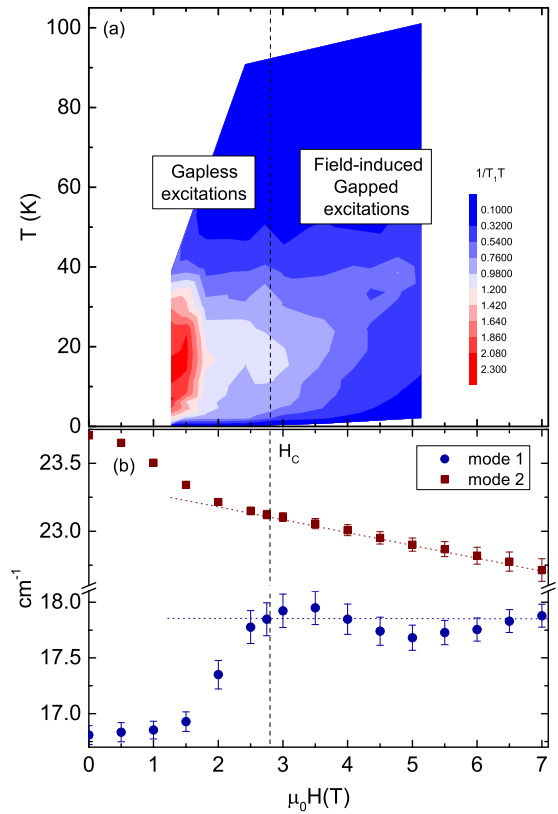}}\par} \caption{\label{fig:phasediagram} (a): Contour plot of $1/T_1T$ vs temperature in different magnetic fields. (b): Field dependencies of the magnon modes at 2\,K estimated from Terahertz spectroscopy. The dotted lines guide the eye.} 
\end{figure}

The behavior of $1/T_1T$ as a function of temperature changes when crossing the field of 2.8\,T. In fields below 2.8\,T, $1/T_1T$ shows a broad hump centered around 20\,K, well below $T_N$, whereas no clear anomaly is observed at $T_N\simeq 38$\,K. This is quite unusual, as magnetic ordering in conventional antiferromagnets is manifested by a sharp peak in $1/T_1$ and, consequently, in $1/T_1T$ at $T_N$~\cite{nath2009,ranjith2015}. Even though powder samples will show a distribution of transition temperatures, owing to the different evolution of $T_N$ for different field directions, this distribution is by far insufficient to explain the hump feature in $1/T_1T$. For example, from our previous thermodynamic measurements~\cite{majumder2019} we expect that at 2\,T the powder sample shows $T_N=34-38$\,K, whereas the hump in $1/T_1T$ measured in fields below 2\,T is centered around 20\,K and appears to be much broader than 4\,K, the distribution of $T_N$. Therefore, we are led to ascribe the broad hump to low-energy spin excitations, similar to the frustrated triangular antiferromagnet NaCrO$_2$~\cite{Olariu2006} with an unconventional spin dynamics (more clearly visible in the contour plot of Fig.~\ref{fig:phasediagram}(a)). 


\begin{figure}
{\centering {\includegraphics[width=8.5cm]{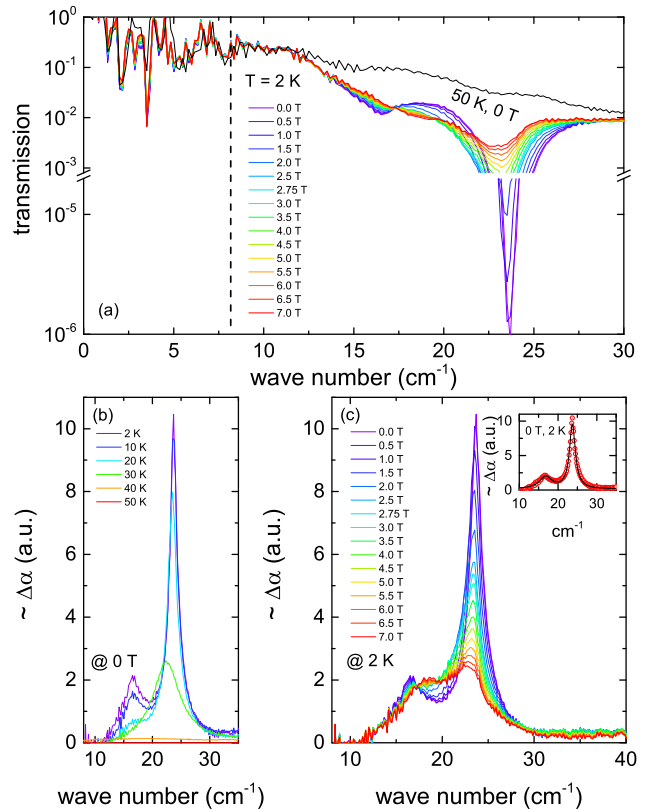}}\par} \caption{\label{fig:THZ}(a): Magnetic field dependent Terahertz transmission raw data. Spectral limitations occur for frequencies below approximately 8\,cm$^{-1}$ (dashed line). (b): Terahertz absorption spectra at different temperatures at zero  magnetic  field. (c): Field  dependence of the Terahertz spectra at 2 K. Inset: the fit of the spectra by considering two Lorentzian lines at zero field and 2 K.} 
\end{figure}

Above 2.8\,T, the behavior of $1/T_1T$ becomes more conventional. It shows a relatively sharp peak around 40\,K reflecting the phase transition for $H\!\!\perp\!b$ and the crossover expected for $H\,\|\,b$~\cite{ruiz2017,majumder2019}.

The Arrhenius representation of $1/T_1$ as a function of $1/T$ resembles an activated behavior in the narrow temperature range of $2-5$\,K, see Fig.~\ref{fig:T1T}(b). However, the data collected below 2\,K suggest that this activated behavior holds only in fields above $H_c$, whereas below $H_c$ no Arrhenius behavior is observed down to 100\,mK. A key finding of our study is that in lower fields the decrease of $1/T_1$ is slower than exponential, suggesting a vanishingly small excitation gap in $\beta$-Li$_2$IrO$_3$ below $H_c$. Using the 1.27\,T data below 500\,mK, we can put $\Delta\simeq 0.4$\,K (34\,$\mu$eV) as the upper limit of the possible excitation gap (Fig.~\ref{fig:T1T}b). This value is well below 1\,\% of the leading exchange couplings in $\beta$-Li$_2$IrO$_3$~\cite{katukuri2016,majumder2019}.

The crossover between the gapless and gapped behavior across $H_c$ can also be deduced from the fact that below 2.8\,T, $1/T_1T$ tends to remain finite at very low temperatures. The values of $1/T_1T$ at the lowest temperature ($\sim$100\,mK) decrease only weakly with increasing field, whereas above $H_c$, $1/T_1T$ shows a sharp drop. The fit of the 3.62\,T data with the activated behavior $1/T_1 = A \times e^{-\Delta/T}$ returns $\Delta\simeq 4$\,K.

The evolution of $1/T_1$ across $H_c$ is clearly different from the one expected for paramagnetic impurities that would show finite $1/T_1T$ even in the applied fields of $3-4$\,T~\cite{kaps2001,zong2008}. On the other hand, it is more difficult to distinguish whether this gap opening occurs for $H\!\!\perp\!b$ or $H\,\|\,b$, as both regimes are probed in the case of the powder sample. The combined nature of $1/T_1$ is confirmed by the fact that the stretching exponent $\beta$ deviates from 1.0 below 40\,K (Fig.~\ref{fig:T1T}a) and indicates the presence of multiple relaxation processes~\cite{johnston2006}, whereas previous single-crystal measurements~\cite{majumder2019} revealed simple exponential behavior ($\beta=1.0$) at all temperatures when the field is applied along $b$ or perpendicular to $b$. Since changes in the magnetic ground state occur for $H\,\|\,b$ only, it is plausible to associate the gap opening with the transformation between the incommensurate and commensurate states. However, a simultaneous gap opening in the incommensurate state for $H\!\!\perp\!b$ can not be excluded.

\textit{THz spectroscopy.} The spectra were measured in the frequency range up to 80\,cm$^{-1}$ on a 3\,mm-thick pressed pellet (see Fig.~\ref{fig:THZ}). Interference fringes prevent us from quantitatively analyzing the data below 8\,cm$^{-1}$, although no appreciable temperature and field dependence was observed below this frequency as seen in the raw transmission spectra in Fig.~\ref{fig:THZ}(a). Due to this spectral limitation our THz data do not provide information about the lowest-energy magnetic modes, responsible for the lack of a spin gap below $H_c$ and for the small spin gap ($\Delta$ = 4\,K = 3\,cm$^{-1}$) right above $H_c$, as observed by NMR. On the other hand, at higher frequencies, there are two sharp resonances emerging in the THz spectra, centered around 17\, and 24\,cm$^{-1}$ in zero field (Fig.~\ref{fig:THZ} (b)). These features are due to zone-center magnons, as they appear only below $T_N$ and remain relatively narrow compared to the excitation continua probed by Raman spectroscopy~\cite{glamazda2016}. The frequencies of 17 and 24\,cm$^{-1}$ correspond to temperatures of about 25 and 35\,K, respectively, and may account for the broad hump in $1/T_1T$ observed in the same temperature range. 

The intensity of the higher-frequency mode is strongly suppressed by the applied magnetic field, whereas the lower-frequency mode retains its intensity and shifts toward higher frequencies, as displayed in Fig.~\ref{fig:THZ}(c). By fitting the spectra with a combination of Lorentzians, we track the field evolution of both peaks, as shown in Fig.~\ref{fig:phasediagram}(b). Both modes show anomalies with respect to their field-dependent resonance frequencies at $H_c$. Moreover, the lower-energy mode becomes field-independent above $H_c$. As our measurements are performed on a polycrystalline sample, zero-field incommensurate order survives in those crystallites where $H\!\!\perp\!b$, but other crystallites should gradually develop commensurate order above $H_c$. Therefore, it seems plausible that the lower-frequency mode occurs in both incommensurate and commensurate states, thus having a nearly constant intensity, whereas the higher-frequency mode belongs to the incommensurate state only. 

\textit{Discussion and Summary.} Long-range order in frustrated magnets manifests itself by low-lying magnon-like excitations, and $\beta$-Li$_2$IrO$_3$ is no exception. Interestingly, these magnons are not gapped, unlike in $\alpha$-RuCl$_3$, where the zero-field magnon gap of 0.4\,meV ($\approx$ 4.8\,K) has been reported~\cite{ran2017}. Magnon gaps of planar iridates, Na$_2$IrO$_3$ and $\alpha$-Li$_2$IrO$_3$, remain to be measured, but tentative inelastic neutron scattering studies suggest that these gaps (if any) should be smaller than 2\,meV ($\approx$ 24\,K) in Na$_2$IrO$_3$~\cite{choi2012} and 1\,meV ($\approx$ 12\,K) in $\alpha$-Li$_2$IrO$_3$~\cite{choi2019}. 

The size of the magnon gap may serve as a useful experimental constraint for the microscopic parametrization of $\beta$-Li$_2$IrO$_3$ within the framework of the $JK\Gamma$ model~\cite{kim2015,lee2016,stavropoulos2018}, where $J$, $K$, and $\Gamma$ stand for the isotropic (Heisenberg) exchange, Kitaev exchange, and off-diagonal anisotropy, respectively. According to linear spin-wave calculations by Ducatman \textit{et al.}~\cite{ducatman2018}, gapless spectra are expected along the lines $J=0$ and $K=\Gamma$, but only in the latter case the gapless state is caused by symmetry, and the gap vanishes identically. In contrast, for $J=0$ the gap arises from higher-order corrections and may be small but non-zero. Our data put a very low upper limit of 34\,$\mu$eV on the gap size and are thus fully compatible with the $K=\Gamma$ regime that is also consistent with the \textit{ab initio} results for $\beta$-Li$_2$IrO$_3$~\cite{majumder2018}. However, the $J\simeq 0$ and $K\neq\Gamma$ regime can not be completely excluded either, because $J$ is indeed small, around 4\,K, according to the numerical analysis of the field-induced transformation around $H_c$~\cite{rousochatzakis2018}.

With $K\simeq\Gamma$ and small $J$ as the most plausible parametrization, linear spin-wave theory~\cite{ducatman2018} predicts the lowest optical magnon mode at $\omega_{\mathbf q=0}/\sqrt{J^2+K^2+\Gamma^2}=0.30\!\div\!0.35$ that corresponds to $21\!\div\!32$\,cm$^{-1}$ for $\sqrt{J^2+K^2+\Gamma^2}=100\!\div\!150$\,K in reasonable agreement with the experimental frequencies. The two magnon modes seen experimentally (Fig.~\ref{fig:THZ}) may be due to the fact that the spin-wave calculations are done for the commensurate approximant, while the real magnetic structure is incommensurate. Spin-wave theory does not predict any further zone-center magnons up to $\omega_{\mathbf q=0}/\sqrt{J^2+K^2+\Gamma^2}\simeq 0.9$, which is at the upper limit or even goes beyond the frequency range of our measurement.

Whereas the incommensurately ordered state of $\beta$-Li$_2$IrO$_3$ is gapless in zero field, the field-induced state appears to be gapped, similar to the quantum paramagnetic state of $\alpha$-RuCl$_3$~\cite{baek2017,hentrich2018}. Above $H_c$, $\beta$-Li$_2$IrO$_3$ reveals at least one sharp excitation with $q=0$ and no excitation continuum within the frequency range of our study, which is different from the field-induced state of $\alpha$-RuCl$_3$ characterized by the coexistence of broad and narrow $q=0$ modes~\cite{wang2017,little2017,sahasrabudhe2020}.

In summary, we used $^7$Li NMR to demonstrate the vanishingly small magnon gap in the incommensurately ordered state of $\beta$-Li$_2$IrO$_3$ and a linear increase in the magnon gap of the field-induced commensurate state above $H_c$. Two $q=0$ excitations are observed by THz spectroscropy and may be responsible for the unusual hump observed in $1/T_1T$ below the magnetic transition. All observed excitations resemble magnons and appear below $T_N$ only. No broad continuum-like features are detected up to 80\,cm$^{-1}$, in contrast to $\alpha$-RuCl$_3$. 

\acknowledgments
MM and AT acknowledge Ioannis Rousochatzakis and Natalia Perkins for fruitful discussions. The work in Augsburg was supported by the German Research Foundation (DFG) via the Project No. 107745057 (TRR80) and by the Federal Ministry of Eduction and Research through the Sofja Kovalevskaya Award of Alexander von Humboldt Foundation.

\bibliography{li2iro3-beta}

\end{document}